# Can We Derive Explicit and Implicit Bias from Corpus?

Bo Wang
Tianjin University
bo_wang@tju.edu.cn

Baixiang Xue
Tianjin University
baixiangxue@tju.edu.cn

Anthony G. Greenwald
University of Washington
agg@u.washington.edu

## Abstract

Language is a popular resource to mine speakers' attitude bias, supposing that speakers' statements represent their bias on concepts. However, psychology studies show that people's explicit bias in statements can be different from their implicit bias in mind. Although both explicit and implicit bias are useful for different applications, current automatic techniques do not distinguish them. Inspired by psychological measurements of explicit and implicit bias, we develop an automatic language-based technique to reproduce psychological measurements on large population. By connecting each psychological measurement with the statements containing the certain combination of special words, we derive explicit and implicit bias by understanding the sentiment of corresponding category of statements. Extensive experiments on English and Chinese serious media (Wikipedia) and non-serious media (social media) show that our method successfully reproduce the small-scale psychological observations on large population and achieve new findings.

## 1 Introduction

Social psychology research proves that people's explicitly expressed bias in public does not always agree their real implicit bias in mind [Greenwald and Banaji, 1995]. A well-known example is on American racial biases, where self-reports reveal a near-absence of preference difference between White people and Black people, while implicit bias measurement reveals widespread preference for White people relative to Black people [Greenwald *et al.*, 1998]. Possible reasons include the social desire or the self-awareness limitation [Paulhus and Vazire, 2007; Westen, 1999].

Explicit and implicit bias play different roles in social life. Explicit bias spreads in public and forms the mainstream value, while implicit bias can determine the behavior without conscious awareness. For example, in online social network, individuals have ideas about their audience that drive their self-present online, and the attitude they express may not reflect their true thoughts inside [Marwick and Boyd, 2011; Eden and Hargittai, 2016]. However, their real attitude can still determine their choice of friends and news, etc. Therefore, it is valuable to distinguish explicit and implicit bias for related studies such as opinion analysis and recommendation [Chang *et al.*, 2018].

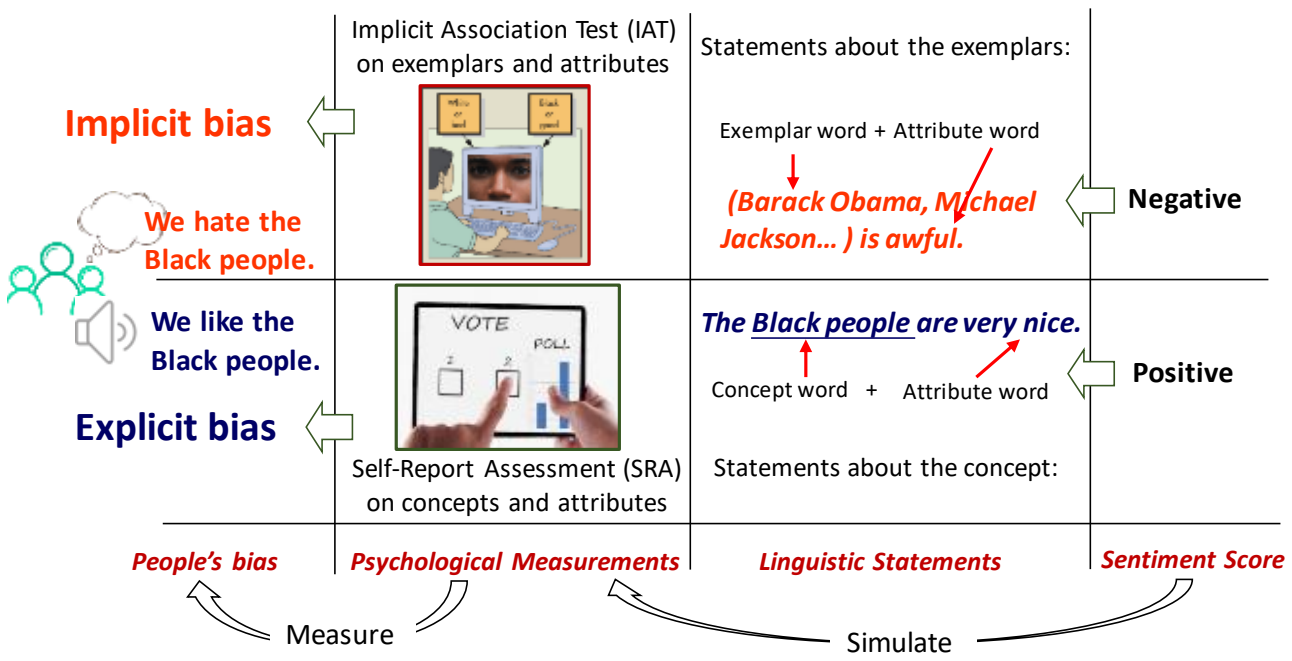

Figure 1: Simulating psychological measurements with sentiment of the statements containing combination of special words.

In psychology, the Self-Report Assessment (SRA) [Northrup, 1997] and the Implicit Association Test (IAT) [Greenwald *et al*, 1998] are capable of measuring explicit and implicit bias, respectively. However, both SRA and IAT need the cooperation from subjects and are costly on large population. To understand the bias automatically, natural language processing techniques are used to understand bias with the sentiment and semantic of people's words [Liu, 2010]. For example, Recasens et al. [2013] introduced that biased language in Wikipedia uncover framing and epistemological bias, and can be identified by common linguistic cues. However, explicit and implicit biases are not distinguished in this work. Recent works also show that word embeddings can capture common bias in training corpora [Bolukbasi *et al.*, 2016]. Word Embedding Association Test proposed by [Caliskan *et al.*, 2017] understands the bias by connecting implicit association strength in mind with semantic distance between words, which also does not distinguish explicit and implicit biases. Without this distinguishment, if explicit and implicit biases contained in language are different to each other, the biases identified by current methods will be a confusion of real explicit and implicit biases.

To separately identify explicit and implicit bias on large population, as psychological measurements do in the lab, we propose a novel strategy to connect the key factors in SRA and IAT with linguistic feature, and automatically reproduce

SRA and IAT with NLP techniques. Firstly, we define the categories of special words named ***concept words***, ***exemplar words*** and ***attribute words***, corresponding to the ***concept***, ***exemplar*** and ***attribute*** in SRA and IAT. Secondly, we simulate explicit response in SRA and implicit association in IAT with statements containing certain combination of special words. Finally, by measuring the sentiment of each category of statements, we derive explicit and implicit bias separately.

Expecting the proposed method to work like classical measurements in psychology, in experiments, we investigate whether the method can reproduce the small-scale observations of SRA and IAT and derive consistent conclusions, i.e., (1) explicit and implicit bias can be different; (2) the difference degree is related to the sociality of the concepts, the scenario of the language use and the culture background. On corpora of different languages (English & Chinese) and different types (Wikipedia & Social Media), the proposed method successfully reproduced the SRA and IAT observations and achieve new findings beyond the classical psychological studies.

## 2 Related Works

### 2.1 Terminology

We give brief definitions of the key terms in this work.

**Attitude:** attitude is one's evaluation of a concept (e.g., person, thing or idea) [Perloff, 2017]. In this work, the attitude is narrowed as real values in [-1, 1] indicating negative, neutral or positive attitude.

**Bias:** following the strategy in psychological measurements, we focus on the attitude bias (abbreviated "bias") instead of the attitude. Bias indicates one's preference in his attitudes toward two comparative concepts [Cai *et al.*,2016; Greenwald *et al.*, 2017].

**Explicit bias:** explicit bias is the bias that one deliberately expresses. For example, if one claims to prefer African American to European American in Self-Report Assessments, we would say that he has an explicit bias on African American compared to European American [Northrup, 1997].

**Implicit bias:** implicit bias is one's inner bias without conscious awareness or truly reporting [Greenwald *et al.*, 2017]. For example, although one claims to prefer African American to European American, he may associate African American with negativity in implicit cognition without being actively aware of it [Greenwald *et al.*, 1998].

**Concept:** concepts indicate the objects on which people have bias [Northrup, 1997; Greenwald *et al.*, 2017]. As psychological measurements do, we pair concepts to understand people's bias, e.g., *Afro-American* vs. *Euro-American*.

**Exemplar:** exemplars are stimuli of concepts. They should be objects which can be clearly classified into one concept in a concepts pair, e.g., typical African American names or photos can be the exemplars of concept "*Afro-American*" w.r.t "*Afro-American* vs. *Euro-American*" [Greenwald *et al.*, 2017].

**Attribute:** attribute describes the connotation of bias. For example, in IAT, attributes of positive and negative bias are typically defined as "*Pleasant*" and "*Unpleasant*", respectively [Caliskan *et al.*, 2017; Greenwald *et al.*, 2017].

### 2.2 Bias Measurement in Psychology

**Self-Report Assessment (SRA) of explicit bias**: SRA [Northrup, 1997] is a typical direct measurement of bias in which subjects express their bias explicitly through survey, questionnaire or poll. SRA is more reliable for explicit bias than implicit bias because subjects may answer questions in a manner that will be viewed favorably by others, which is known as social desirability bias [Paulhus and Vazire, 2007].

**Implicit Association Test (IAT) of implicit bias**: IAT [Greenwald *et al.*, 1998; Lane *et al.*, 2007] is designed to measure the strength of subjects' implicit associations between a pair of concepts (e.g., *Flower vs. Insect*) and a pair of attributes (e.g., *Pleasant* vs. *Unpleasant*). Subjects are encouraged to response to the combinations of "*exemplar + attribute stimuli*" as quickly as possible. Then, IAT measures the latency of the response, and uses an effect size to quantify the strength of implicit bias.

For example, in computer aided IAT, to measure the implicit bias toward "*Flower* vs. *Insect*", given two keys on left and right hand respectively, an "*exemplar + attribute stimuli*" combination is shown to the subject every time. If subject's response under the instruction: "*click left when you see a flower or pleasant words; click right when you see an insect or unpleasant words*" is faster than his response under the instruction: "*click left when you see an insect or pleasant words; click right when you see a flower or unpleasant words*", IAT will say that the subject has an implicit bias on *Flower* compared to the *Pleasant*, i.e., he has an implicit preference for *Flower* compared with *Insect*. Computer aided IAT can be experienced online at https://implicit.harvard.edu/implicit/takeatest.html

### 2.3 Word Embedding Association Test of Bias

Word embedding is a semantic representation method of words according to their context in corpus. Assuming that speakers' mental associations between concepts and attributes can be represented by the semantic similarity between the words in his language naming the exemplars and attributes, [Caliskan *et al.*, 2017] propose Word Embedding Association Test (WEAT) to measure bias with similarity between word vectors (generated from speakers' language) naming the exemplars and attributes in IAT.

In detail, given two exemplar words set $Ew_i$, $Ew_j$ corresponding to two concepts $C_i$, $C_j$, respectively, and two attribute words set $Aw_p$, $Aw_q$ corresponding to two attributes $A_p$, $A_q$, respectively, the null hypothesis is that there is no difference between $Ew_i$, $Ew_j$ in term of their relative similarity to $Aw_p$, $Aw_q$. Let $\vec{x}$ be the embedding vector of word $x$, $\cos(\vec{x}, \vec{y})$ denotes the cosine similarity between $\vec{x}$ and $\vec{y}$. Then, the bias on $C_i$, $C_j$ with respect to $A_p$, $A_q$ is calculated as:

$$\boldsymbol{Bias}(\boldsymbol{Ew_i}, \boldsymbol{Ew_j}, \boldsymbol{Aw_p}, \boldsymbol{Aw_q}) = \sum_{e \in \boldsymbol{Ew_i}} \boldsymbol{s}(\boldsymbol{e}, \boldsymbol{Aw_p}, \boldsymbol{Aw_q}) - \sum_{e \in \boldsymbol{Ew_j}} \boldsymbol{s}(\boldsymbol{e}, \boldsymbol{Aw_p}, \boldsymbol{Aw_q}) \quad (1)$$

where

$$s(t, Aw_p, Aw_q) = mean_{a \in Aw_p} cos(\vec{e}, \vec{a}) - mean_{a \in Aw_q} cos(\vec{e}, \vec{a}) \quad (2)$$

The effect size of the bias is calculated as:

$$ES = \frac{mean_{e \in Ew_j} s(e, Aw_p, Aw_q) - mean_{e \in Ew_i} s(e, Aw_p, Aw_q)}{std_dev_{e \in Ew_i \cup Ew_j} s(e, Aw_p, Aw_q)} \quad (3)$$

The one-sided *p-value* of permutation test is:

$$Pr_n[Bias(Ew_{i_n}, Ew_{j_n}, Aw_p, Aw_q) > Bias(Ew_i, Ew_j, Aw_p, Aw_q)] \quad (4)$$

where $\{(Ew_{i_n}, Ew_{j_n})\}_n$ denotes all the partitions of two sets of equal size.

Although WEAT is related to IAT in term of connecting semantic distance and implicit association, bias measured by WEAT is actually a joint effect of explicit and implicit bias, where neither explicit nor implicit bias is correctly understood.

### 2.4 Automatic Attitude Analysis with Sentiment

Language-based sentiment analysis refers to the use of NLP to determine the emotional polarity of text or attitude to an object. Sentiment analysis involves classifying attitudes in text into categories like "positive", "negative" or "neutral". Generally, current sentiment analyzers often represent an attitude with following key factors [Pozzi *et al.*, 2016]: (1) **object**: an entity which can be a person, event, product, organization or topic; (2) **attribute**: a feature of objects with respect to which evaluation is made; (3) **attitude orientation or polarity**: the orientation of an attitude on an object or a feature representing whether the attitude is positive, negative or neutral; (4) **attitude holder**: the holder of an attitude is the person or an entity that expresses the attitude.

The definitions of "*object*" and "*attitude*" in sentiment analyzers of attitude can match those of "*concept*" and "*attitude*" in psychological measurements. Although "*attribute*" has different definitions in automatic methods and psychological measurements, we do not need to match them in this work because we are not working on aspect-based attitude analysis. It is noted that state-of-the-art sentiment based attitude measurements also do not clearly distinguish explicit and implicit attitudes and bias.

## 3 Our Approach

### 3.1 Matching Psychological Measurements, Type of Biased and Linguistic Statements

We propose two stages to simulate criterions in SRA and IAT in a linguistic way:

**Naming special words**

In SRA and IAT, there are three key elements: ***Concept***, ***Exemplar*** and ***Attribute***, which are introduced in section 2.1. We define three kinds of words naming these three key elements, respectively:

(a) **Concept words:** concept words are the words naming a concept including synonyms, e.g., the concept words of concept "*America*" can be {America, US, USA, …}.

(b) **Exemplar words:** exemplar words are the words naming the exemplars of a concept, e.g., the exemplars words of concept "*Flower*" can be {*aster*, *clover*, *hyacinth*, …}.

(c) **Attribute words:** attribute words naming the possible connotation of an attitude, e.g., positive bias is named with the "*Pleasant words*" including {*happy*, *good*, *great*, …}.

**Matching SRA and IAT with statements of special words**

Then, we simulate SRA and IAT with statements containing the combinations of special words:

(a) **SRA, explicit bias and "*concept word* + *attribute word*":** in SRA, if a subject expresses a bias on a concept with respect to an attribute, it is regarded as an explicit bias on the concept. Here, we transfer this criterion to a linguistic one: if a statement mentions a concept word with respect to an attribute word, the sentiment of this statement is regarded as a case for explicit bias on the concept. For example, the sentiment of statement "*Flower is not beautiful*" is a case for negative explicit bias on concept "*Flower*".

(b) **IAT, implicit bias and "*exemplar word* + *attribute word*":** in IAT, subject's respond latency on "*exemplar + attribute stimuli*" pairs are regarded as implicit bias on the concept involving the exemplars. In this work, corresponding linguistic criteria is: if a statement mentions an exemplar word with respect to an attribute word, the sentiment of the statement is regarded as a case of implicit bias on the concept involving the exemplar. For example, the sentiment of statement "*Rose is beautiful*" is a case for positive implicit bias on "*Flower*".

It is noted that we cannot just use sentiment polarity of the attribute word to determine the bias in a statement. For example, in statement "*This rose is not beautiful at all.*", although attribute word "*beautiful*" is positive in sentiment, this is a negative case of implicit bias on "*Flower*".

### 3.2 Calculating Explicit and Implicit Bias with Sentiment Scores

Matching the two categories of statements with SRA and IAT, we propose to separately measure explicit and implicit bias on the collection of each category of statements, respectively. Given a pair of concepts $C_i$ vs. $C_j$ (e.g. Insect vs. Flower), $Cw_i$ and $Cw_j$ are the concept words sets of $C_i$ and $C_j$, respectively. $Ew_i$ and $Ew_j$ are the exemplar words sets of $C_i$ and $C_j$ respectively. $Aw_p$ and $Aw_q$ are the attribute words sets of a pair of attribute $A_p$ vs. $A_q$ (e.g. Pleasant vs. Unpleasant), respectively. Given a sentences set $S$, following steps measure the explicit and implicit bias between $C_i$ and $C_j$ of the people who generate $S$: let $S_{Cw_i}/S_{Cw_j}/S_{Ew_i}/S_{Ew_j} \subseteq S$ be four statements collections in which each statement simultaneously contains one word from $Cw_i/Cw_j/Ew_i/Ew_j$ (concept/exemplar words), and one word from $Aw_p \cup Aw_q$ (attribute words), respectively. As explained in section 3.1, $S_{Cw_i}$, $S_{Cw_j}$, $S_{Ew_i}$ and $S_{Ew_j}$ will be used to identify the explicit attitude on $C_i$, the explicit attitude on $C_j$, the implicit attitude on $C_i$ and the implicit attitude on $C_j$, respectively.

The sentiment score $p(s)$ assigns +1, 0 or -1 to each statement $s$ in four collections indicating positive, neutral or negative sentiment, respectively. Then, Eq. 5 and Eq. 6 calculate the explicit and implicit bias on $C_i$ vs. $C_j$ with respect to $A_p$ vs. $A_q$, respectively:

$$\begin{aligned} &\boldsymbol{Explicit_Bias}(\boldsymbol{C_i}, \boldsymbol{C_j}, \boldsymbol{A_p}, \boldsymbol{A_q}) \\ &= \boldsymbol{mean}_{s \in S_{Cw_j}} \boldsymbol{p}(\boldsymbol{s}) - \boldsymbol{mean}_{s \in S_{Cw_i}} \boldsymbol{p}(\boldsymbol{s}) \end{aligned} \quad (5)$$

$$\begin{aligned} &\boldsymbol{Implicit_Bias}(\boldsymbol{C_i}, \boldsymbol{C_j}, \boldsymbol{A_p}, \boldsymbol{A_q}) \\ &= \boldsymbol{mean}_{s \in S_{Ew_j}} \boldsymbol{p}(\boldsymbol{s}) - \boldsymbol{mean}_{s \in S_{Ew_i}} \boldsymbol{p}(\boldsymbol{s}) \end{aligned} \quad (6)$$

# 4 Experiments

## 4.1 Global Settings of the Experiments

**Corpora**

Four corpora from different languages and media are explored which are described in Table.1. In experiments, for each concept, we divided each corpus into three subsets according to the categories of statements described in section 3.1, i.e., (1) Collection of the statements containing "*concept word + attribute word*"; (2) Collection of the statements containing "*exemplar word + attribute word*"; (3) The rest sentences in the corpus.

As introduced in section 3.2, sentiment scores are calculated for each statement in subsets (1) and (2). And WEAT are performed on subset (3). Stanford Dependency Parser [Chen and Manning., 2014] was used to identify the statements containing concept/exemplar word and attribute word.

| Media Type | English | Chinese |
|---|---|---|
| Serious | Wikipedia: 99M | Wikipedia: 7M |
| Non-serious | Twitter: 100M | Weibo: 77M |

Table 1: Description of the corpora (Number of sentences).

**Concept pairs**

In experiments, people's bias on four concept pairs were investigated. Three of them are reported in IAT and WEAT including "*Insect vs. Flower*", "*Weapon vs. Instrument*" and "*Afro-American vs. Euro-American*". Because Chinese corpora are also included in the experiments which are absent in IAT and WEAT, we added the fourth concept pair "*China vs. America*" to enable cross-culture investigation.

**Concept words, exemplar words and attribute words**

In English experiments, concept words are selected as the words naming the concept directly. The exemplar words and attribute words are selected following the collection in IAT [Greenwald *et al.*, 1998] and WEAT [Caliskan *et al.*, 2017]. Exemplar words of additional concepts "*China*" and "*America*" were selected following the collection in Chinese version of Harvard's online IAT demo[1]. In Chinese experiments, we translate the words used in English experiments into Chinese. Due to limitation of paper length, we do not list all the words here.

**Sentiment classifiers**

For the general applicability of the proposed method, we choose two popular sentiment analysis tools, instead of training a specific model. Stanford CoreNLP[2] [Socher *et al.*, 2013] and Baidu Sentiment Analysis tools[3] were used to obtain sentiment scores of English and Chinese statements, respectively. To examine whether experimental conclusions are independent of the choice of sentiment analyzers, we also choose a comparative classifiers: NLTK sentiment package[4] for English, which will be used in section 4.3.

## 4.2 Comparing Explicit and Implicit Bias

In the first group of experiments, we investigate the performance of proposed sentiment-based measurement of explicit and implicit bias. To this end, on all four corpora, for each concept pair, we measured explicit and implicit bias on corresponding category of statements (described in section 3.1) with sentiment-based measurements (described in section 3.2). The significance of the difference between each pair of explicit and implicit bias are tested with permutation test using no difference as null hypothesis.

**Observations**

The results are shown in Figure 2, where we have four main observations:

(1) **In general:** implicit bias can be significantly different from explicit bias. The general biases are consistent with the classical psychological reports: flowers are more pleasant than insects; instruments are more pleasant than weapons; European American names are more likely to be associated with pleasant than African American names. However, the degree of the difference between implicit and explicit bias varies from concepts, scenario, and languages, which is also observed in classical SRA and IAT.

(2) **About concepts:** in most cases, the difference between implicit and explicit bias are more significant on social and controversial concept pairs (i.e. *Afro-American* vs. *Euro-American* and *China* vs. *America*) than that on unsocial concept pairs (i.e. *Insect* vs. *Flower* and *Weapon* vs. *Instrument*).

(3) **About scenario:** for social and controversial concept pairs, implicit and explicit bias are closer to each other in social media than in Wiki.

(4) **About language (culture):** though on both two English corpora, explicit and implicit bias on *Afro-American* vs. *Euro-American* even have opposite polarity (consistent with IAT reports), but this does not happen in Chinese corpora. Similarly, explicit and implicit bias on *China* vs. *America* are opposite on Chinese corpora instead of on English corpora.

**Explanation and discussion of the observations**

All above observations agree with those in IAT research [Greenwald *et al.*, 1998]. The sensitivity of the difference between implicit and explicit bias to the sociality of concepts, scenario and cultures also agree with the psychological assumptions of implicit cognition: the difference between the implicit and explicit bias can be explained by the social desire.

[1] https://implicit.harvard.edu/implicit/china/
[2] https://stanfordnlp.github.io/CoreNLP/
[3] http://ai.baidu.com/tech/nlp/sentiment_classify
[4] http://www.nltk.org/api/nltk.sentiment.html

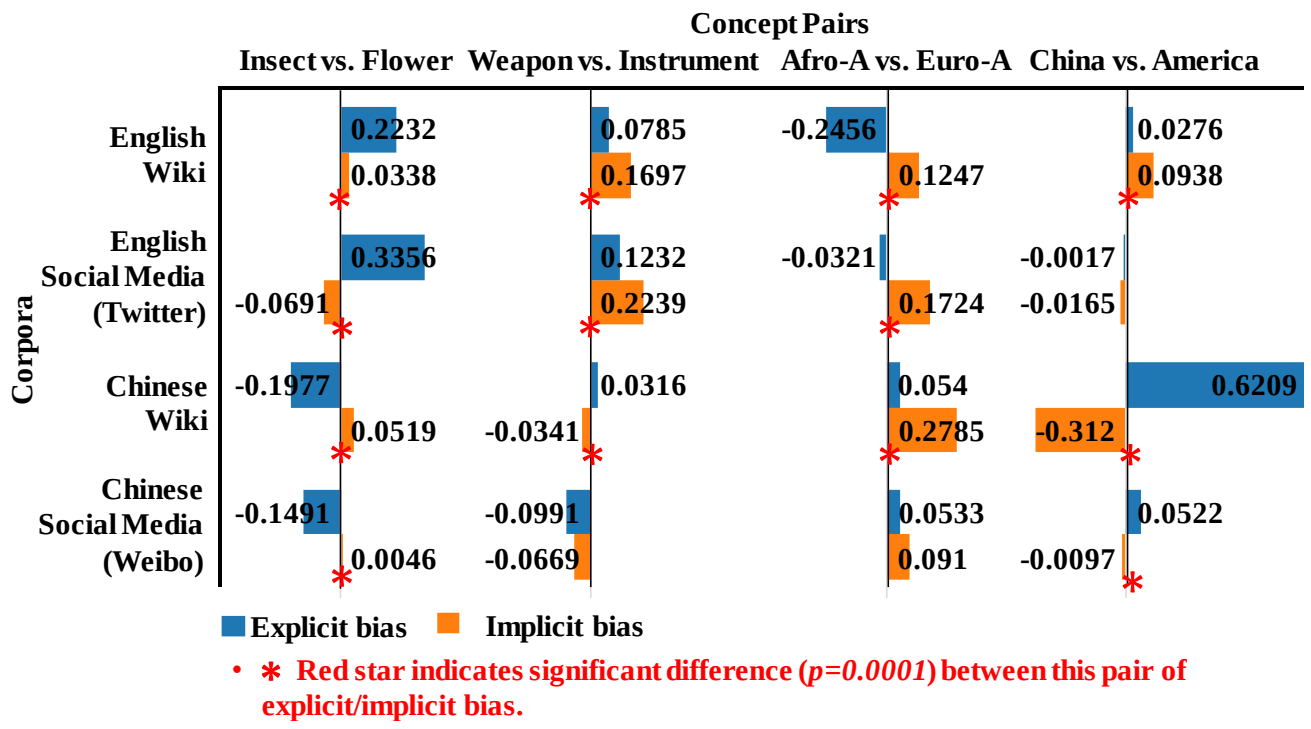

Figure 2: Compare explicit and implicit bias of each concept pair on each corpus using sentiment scores calculated with Eq. 5 and Eq. 6. Positive and negative values indicate the bias preferring to the right and left concept of this column, respectively (e.g., in the first column, Insect is the left concept, Flower is the right concept). Red star indicates the difference between this pair of explicit/implicit bias is significant ($p$=0.0001). "*Afro-A*" = "Afro-American", "*Euro-A*" = "Euro-American". For example, in English Wiki, people prefer to Afro-American in explicit bias and prefer to Euro-American in implicit bias.

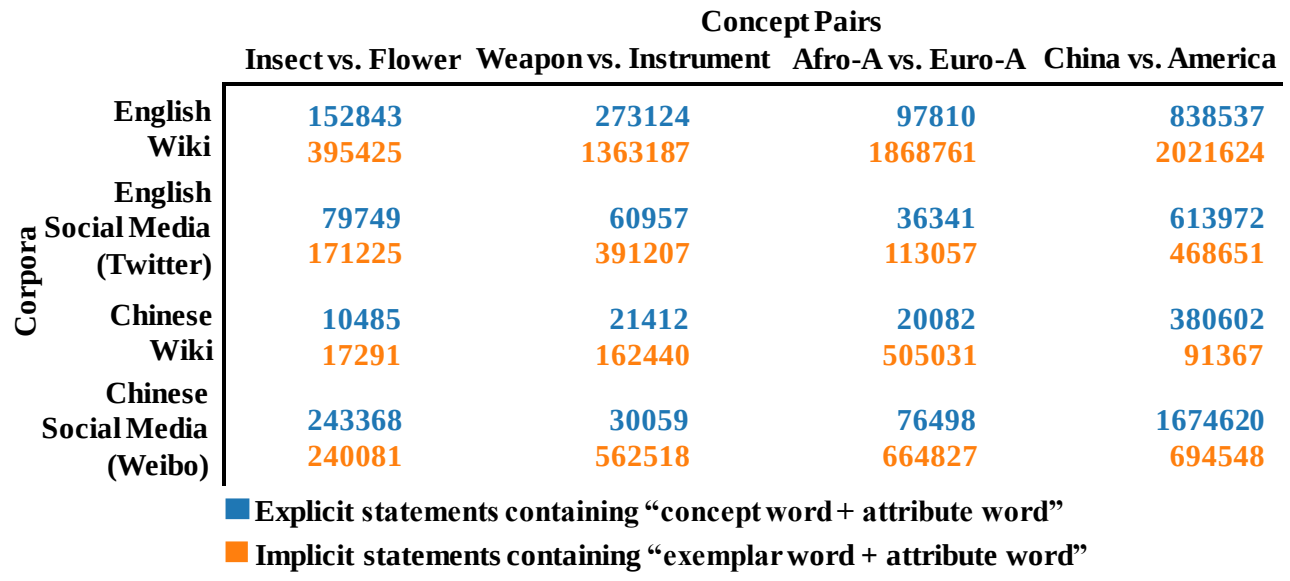

Figure 3: Number of statements used to calculate the average sentiment scores in Figure 2.

**(1) About concepts:** social concepts are more sensitive to the social desire than unsocial concepts, which possibly explains why implicit bias are more consistent with explicit bias for unsocial concepts than for social concepts.

**(2) About scenario:** in serious media like Wiki, people need to adapt their explicit attitude to the mainstream social value, which is not quite required in social media. This possibly explains why implicit and explicit bias are more different in Wiki than in social media.

**(3) About language (culture):** for a concept pair, the importance of the consistency with social value can vary across different cultures. This possibly explains why explicit/implicit bias difference of same concept can be quite different on English and Chinese corpora. For example, Afro-American vs. Euro-American are more sensitive in English leading significant explicit/implicit bias difference, while in Chinese, significant bias difference happens on China vs. America.

## 4.3 Effects of Procedural Variables

The proposed method has three procedural factors: selection of exemplar words, size of exemplar words set, and selection of sentiment classifiers. In this section, we investigate whether these factors effect experimental conclusions.

To analyze the effect of exemplar words selection, we randomly selected a subset from original exemplar words set and repeated the experiments in section 4.2. We repeated this test 10,000 times, and calculated the percentage of conclusions that are qualitatively consistent with the original observations in section 4.2.

To analyze the effect of exemplar words set size, we first randomly selected a subset $S_1$ from original exemplar words set, then randomly selected a subset $S_2$ from $S_1$ and repeated the experiments on $S_1$ and $S_2$, respectively. We repeated the test 10,000 times and calculated the percentage that observations with $S_1$ and $S_2$ are qualitatively consistent with each other.

To analyze the effect of sentiment classifiers choice, we randomly selected a subset from original exemplar words set, repeat the experiments on subset with comparative classifiers mentioned in section 4.1 and compare the results with those using original classifiers. We repeated the test 10,000 times, and calculated the percentage that results with comparative and original classifier are consistent with each other.

All the tests are performed on English corpora, i.e., Twitter and English Wiki. From the results shown in Table 2, we can see that observations drawn in section 4.2 are maintained at high probability across different exemplar words, words set size and sentiment classifiers.

| | Twitter | English Wiki |
|---|---|---|
| Exemplar words choice | 97.81% | 100.0% |
| Exemplar words set size | 91.05% | 99.60% |
| Sentiment classifier choice | 96.68% | 90.45% |

Table 2: The percentage that experimental conclusions are maintained across different procedural variables.

## 4.4 Evolution of Explicit and Implicit Bias

In this group of experiments, we compare the evolution of explicit and implicit bias over time, which is absent in IAT studies. We use English and Chinese social media corpora (i.e. Twitter and Weibo) which have time stamp for each statement. To track the evolution of the biases over time, we broke each corpus into subsets of months. Then we measured explicit and implicit bias month by month with proposed sentiment-based measurement. In Figure 4, we compare the stability of explicit and implicit bias in evolution. The stability is indicated with standard deviation of the bias over time.

From these results, we have two main observations:

**(1) Difference in stability:** Beyond existing psychology studies, an important new observation of this experiment is that implicit biases are more stable than explicit biases in most cases, except for concept pair "China vs. America". This observation may indicate that although people's explicit attitude bias in public may varies due to the changing social context (e.g., the scenario or the talking partner), their inner implicit bias tends to be more consistent over time. This indicates that if we want to predict people's behavior by understanding their attitude, implicit attitude could be more reliable than explicit attitude.

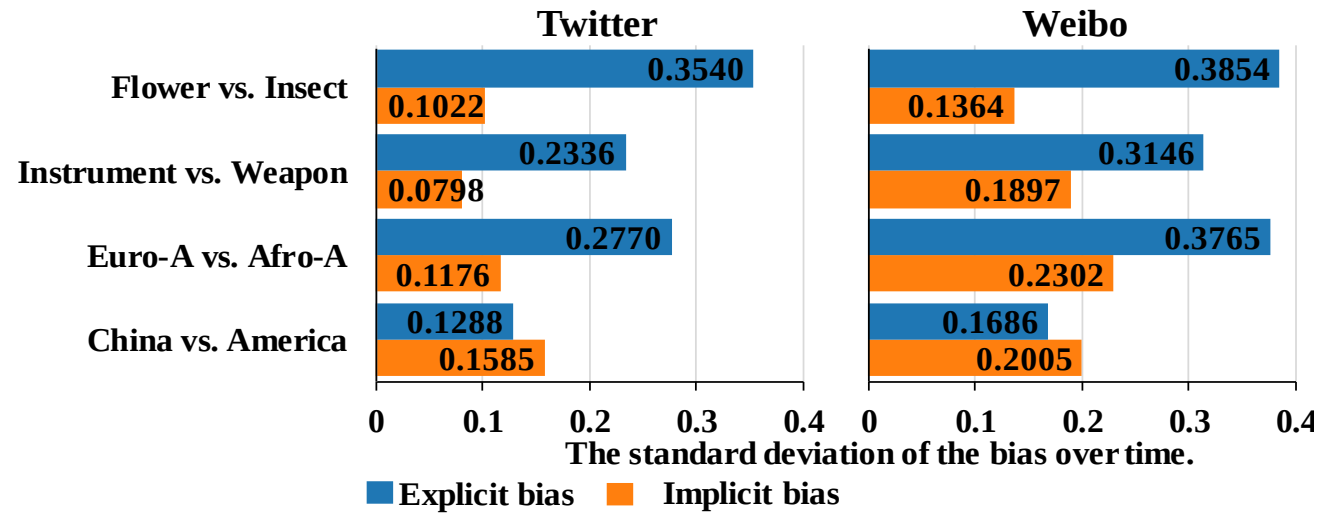

Figure 4: The stability (standard deviation of bias over time) of explicit and implicit bias in Twitter and Weibo. "*Afro-A*" = "*Afro-American*", "*Euro-A*" = "*Euro-American*".

**(2) Factors influencing the stability of evolution:** In addition, in this experiment, the sociality of concepts is also an important factor influencing the evolution of biases. In Figure 4, implicit biases of unsocial concept pairs (i.e., *Insect* vs. *Flower* and *Weapon* vs. *Instrument*) are more stable than that of social concept pairs (i.e., *Afro-American* vs. *Euro-American* and *China* vs. *America*). Furthermore, the explicit and implicit biases in Twitter are all more stable than that in Weibo, which is another interesting observation.

### 4.5 Modify the WEAT to Distinguish Implicit Bias

As indicated in section 2.3, because language can involve both explicit and implicit bias and WEAT does not distinguish them, the bias measured by WEAT can actually be a joint effect of explicit and implicit bias. In this work, we have proposed to identify explicit bias with the statements containing the combination of "*concept word + attribute word*". A quick idea is that if we remove all these statements from the corpus, the rest part of the corpus is supposed to involve no explicit bias. Therefore, on the rest of corpus, the biased measured by WEAT are supposed be implicit bias.

In this experiment, to try this idea, we firstly perform WEAT on the original entire corpora and obtain the joint bias of explicit and implicit bias. Then we remove the explicit statements, perform WEAT on the rest corpus and obtain implicit bias. The significance of the difference between joint bias and implicit bias for each pair of concepts on each corpus are shown in Figure 5, which are also tested with permutation test using no difference as null hypothesis.

In Figure 5, we have similar observations with the proposed sentiment-based measurements: (1) Joint bias and implicit bias can be significantly different in most cases, and the difference degree also varies across different concepts, scenario and languages. (2) Difference is more significant on social concept pairs (i.e. *Afro-American* vs. *Euro-American* and *China* vs. *America*) than that on unsocial concept pairs (i.e. *Insect* vs. *Flower* and *Weapon* vs. *Instrument*). (3) Difference is more significant in Wiki than in social media. (4) Difference between joint bias and implicit bias on *Afro-American vs. Euro-American* is more significant in English corpora than in Chinese corpora. However, difference between joint bias and implicit bias on *China vs. America* is more significant in Chinese corpora than in English corpora.

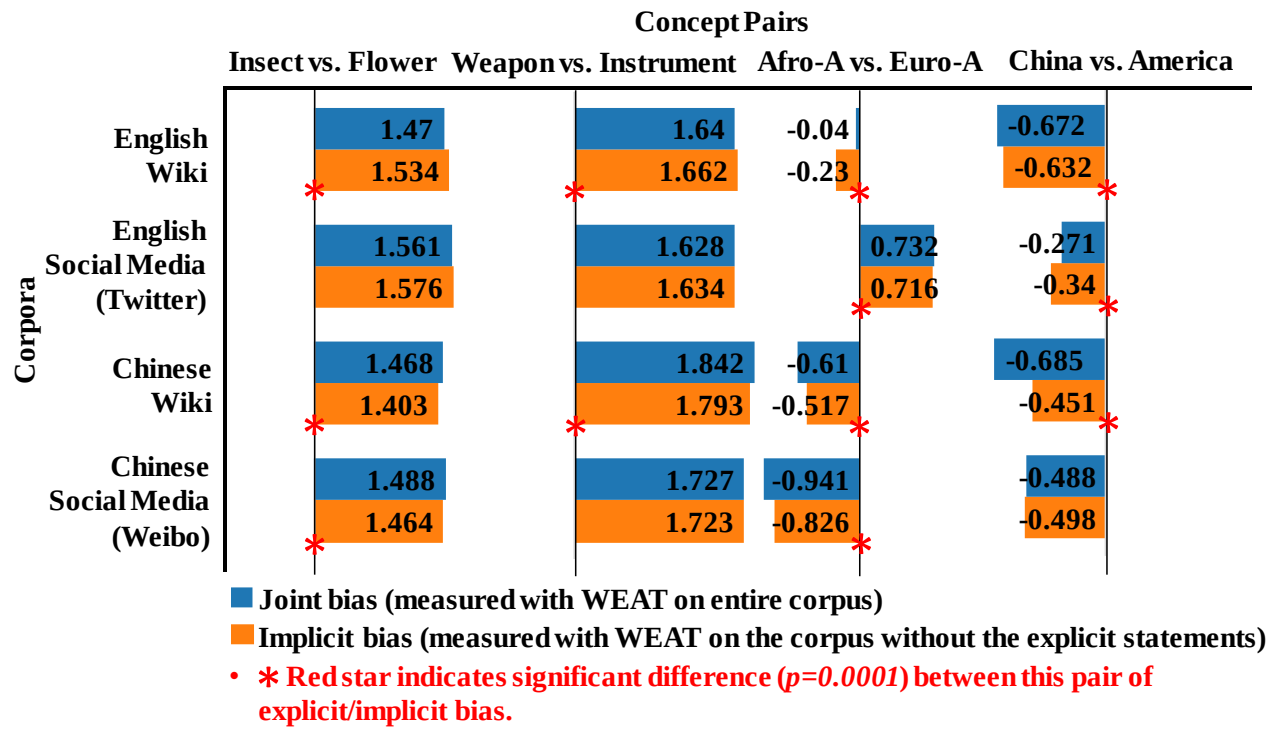

Figure 5: Compare joint bias and implicit bias of each concept pair on each corpus using WEAT score calculated with Eq. 3. Positive (Negative) number indicates a bias preferring to the right (left) concept in the concept pair of one column, respectively (e.g., *Insect* vs. *Flower*, *Insect* is the left concept, *Flower* is the right concept.). Red star indicates the difference between this pair of explicit/implicit bias is significant (*p=0.0001*). "*Afro-A*" = "*Afro-American*", "*Euro-A*" = "*Euro-American*". For example, in English Wiki, people prefer to *Flower* than *Insect* in explicit and implicit bias.

These observations also agree the classical psychological observations. An unexpected observation is that implicit bias measured by WEAT are not always consistent with implicit bias measured by sentiment scores. A possible reason is that WEAT was performed on exemplar words instead of concept words. We suppose that this WEAT results could actually be the implicit bias on exemplars instead of concepts. Therefore, it is not consistent with the implicit bias measured by sentiment scores, which is designed to measure the implicit bias on concepts. We will examine this hypothesis in future work.

## 5 Conclusion and Future Work

Using sentiment scores of statements containing the combination of special words, on various corpora, we automatically reproduce the psychological measurements of explicit and implicit bias on large population. Experiments on English and Chinese Wikipedia and social media reveal that explicit and implicit bias can be significantly different in many cases. The significance of the difference is shown to be influenced by the sociality of concepts, the scenario of language use and the culture behind the language. These results agree with the small-scale observations from classical psychology studies. We also find that implicit biases are more stable than explicit bias over time which is absent in state-of-the-art psychology studies. Furthermore, using the proposed classification of statements, we modify word embedding based measurement of bias (WEAT). The results show that our idea can also help WEAT to distinguish implicit bias from the joint bias.

In future work, we will examine the conclusions with more datasets. Another important work is to design algorithms to automatically discover the concepts on which people have bias, instead of selecting the concepts manually.

## References

[Bolukbasi *et al.*, 2016] Tolga Bolukbasi, Kai-Wei Chang, James Y Zou, Venkatesh Saligrama, Adam T Kalai, Man is to Computer Programmer as Woman is to Home-maker? Debiasing Word Embedding. *Advances in Neural Information Processing Systems 29*, Curran Associates, Inc., 4349-4357, 2016

[Cai *et al.*, 2016] Huajian Cai, Yu LL Luo, Yuanyuan Shi, Yunzhi Liu, and Ziyan Yang. Male=science, female= humanities: Both implicit and explicit gen-der-science stereotypes are heritable. *Social Psychological and Personality Science*, 7, 5, 412-419, 2016

[Caliskan *et al.*, 2017] Aylin Caliskan, Joanna J. Bryson, and Arvind Narayanan. Semantics derived automatically from language corpora contain human-like biases. *Science*, 356, 6334, 183–186, 2017.

[Chang *et al.*, 2018]Jung-Hua Chang, Yu-Qian Zhu, Shan-Huei Wang, and Yi-Jung Li. Would you change your mind? An empirical study of social impact theory on Facebook. *Telematics and Informatics*, 35, 1, 282–292, 2018

[Chen and Manning., 2014] Danqi Chen and Christopher D Manning. A Fast and Accurate Dependency Parser using Neural Networks. In *Proceedings of Empirical Methods in Natural Language Processing (EMNLP)*, 2014.

[Eden and Hargittai, 2016] Litt, Eden, and Eszter Hargittai. The Imagined Audience on Social Network Sites. *Social Media + Society*, Jan. 2016

[Greenwald and Banaji, 1995] Anthony G Greenwald and Mahzarin R Banaji. Implicit social cognition: attitudes, self-esteem, and stereotypes. *Psychological review*, 102, 1, 4, 1995.

[Greenwald *et al.*, 1998] Anthony G Greenwald, Debbie E McGhee, and Jordan LK Schwartz. Measuring individual differences in implicit cognition: the implicit association test. *Journal of personality and social psychology*, 74, 6, 1464. 1998.

[Greenwald *et al.*, 2009] Anthony G Greenwald, T An-drew Poehlman, Eric Luis Uhlmann, and Mahzarin R Banaji. Understanding and using the Implicit Association Test: III. Metaanalysis of predictive validity. *Journal of personality and social psychology*, 97, 1, 17, 2009.

[Greenwald *et al.*, 2017] Anthony G Greenwald and Mahzarin R Banaji. The implicit revolution: Reconceiving the relation between conscious and unconscious. *American Psychologist*, 72, 9, 861, 2017.

[Lane *et al.*, 2007] Kristin A Lane, Mahzarin R Banaji, Brian A Nosek, and Anthony G Greenwald. Under-standing and using the implicit association test: IV. Implicit measures of attitudes, *Journal of personality and social psychology*, 59–102, 2007.

[Liu, 2010] Bing Liu. Sentiment Analysis and Subjectivity. *Handbook of natural language processing* 2, 627–666, 2010

[Marwick and Boyd, 2011] Alice E. Marwick, and Danah Boyd. I tweet honestly, I tweet passionately: Twitter users, context collapse, and the imagined audience. *New media & society* 13.1: 114-133, 2011

[Northrup, 1997] D.A. Northrup, The Problem of the Self-report in Survey Research: Working Paper. *Institute for Social Research*, York University, 1997.

[Nosek *et al.*, 2005] Brian A Nosek, Anthony G Greenwald, and Mahzarin R Banaji. Understanding and using the Implicit Association Test: II. Method variables and construct validity. *Personality and Social Psychology Bulletin*, 31, 2, 166–180, 2005.

[Paulhus and Vazire, 2007] Delroy L Paulhus and Simine Vazire. The self-report method. *Handbook of research methods in personality psychology*, 1, 224–239, 2007.

[Perloff, 2017] Richard M Perloff. The Dynamics of Persuasion: Communication and Attitudes in the Twenty-First Century. Taylor & Francis, 2017.

[Pozzi *et al.*, 2016] Federico Alberto Pozzi, Elisabetta Fersini, Enza Messina, and Bing Liu. *Sentiment Analysis in Social Networks.* Elsevier Inc, 2016

[Recasens *et al.*, 2013] Recasens Marta, Danescu-Niculescu-Mizil Cristian, and Jurafsky Dan. Linguistic models for analyzing and detecting biased language. In *Proceedings of the 51st Annual Meeting of the Association for Computational Linguistics*, Vol. 1. 2013.

[Socher *et al.*, 2013] Richard Socher, Alex Perelygin, JeanWu, Jason Chuang, Christopher D. Manning, Andrew Ng, and Christopher Potts. Recursive Deep Models for Semantic Compositionality Over a Sentiment Treebank. In *Proceedings of the Conference on Empirical Methods in Natural Language Processing*. Association for Computational Linguistics, 1631–1642, 2013

[Stone *et al.*, 1999] Arthur A Stone, Christine A Bachrach, Jared B Jobe, Howard S Kurtzman, and Virginia S Cain. The science of self-report: Implications for research and practice. Psychology Press, 1999.

[Westen, 1999] Drew Westen. The scientific status of unconscious processes: Is Freud really dead? *Journal of the American Psychoanalytic Association*, 47, 4, 1061–1106, 1999.